# Hierarchical structure in phonographic market[1]

## Andrzej Buda

<sup>a</sup>Wydawnictwo Niezalezne, ul. Oriona 15/8 67-200 Głogów, Poland

Abstract

I find a topological arrangement of assets traded in a phonographic market which has associated a meaningful economic taxonomy. I continue using the Minimal Spanning Tree and the Life-time Of Correlations between assets, but now outside the stock markets. This is the first attempt to use these methods on phonographic market where we have artists instead of stocks. The value of an artist is defined by record sales. The graph is obtained starting from the matrix of correlations coefficient computed between the world's most popular 30 artists by considering the synchronous time evolution of the difference of the logarithm of weekly record sales. This method provides the hierarchical structure of phonographic market and information on which music genre is meaningful according to customers.

## 1  Introduction

Phonographic markets are well defined complex systems as old as financial markets. They are studied by economists and sociologists [1-5], but have

---

1  It is strongly recommended to read previously: A. Buda, *Life-time of correlations between stocks on established and emerging markets* in Life-time Of Correlation And Its Application (volume 1), Wydawnictwo Niezależne, Wrocław 2010 never been explored by mathematicians and physicists. Since Thomas Edison (1877) the phonograph, gramophone or record player has been commonly used for playing sound. Although in 1906 these devices were elusive for common people, the Italian tenor Enrico Caruso was the first and only artist who sold more than 1,000,000 copies of his record before The Beatles [7]. In 1948 Columbia released the first ever long playing record (LP) that would hold at least 20 minutes per side [4]. Since 1967's The Beatles 'Sgt Pepper's Lonely Hearts Club Band' long playing records has dominated the phonographic markets worldwide. The new formats (vinyl, cassette, compact disc and mp3, etc.) were more and more popular. Thus, Michael Jackson's 'Thriller' has become the most popular record ever and sold over 110,000,000 copies [6].

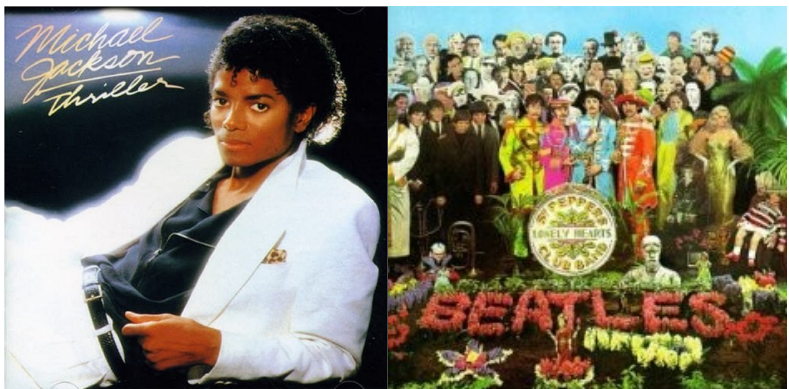

**Pic 1** The most popular record ever – Thriller by Michael Jackson and The Beatles' Sgt Pepper's Lonely Hearts Club Band - the 15[th] best selling record of all time.

On the other hand, currently his success is impossible to repeat because of digital piracy, marginal utility, etc. [8]. Global music sales in 2009 fell by 7% to US $17 billion. This is disappointing, but amid the decline there are some very positive points. No fewer than thirteen countries saw music sales grow in 2009, including important markets such as Australia, Brazil, South Korea, Sweden and the UK [6]. Digital sales in some of those markets rose at very encouraging rates, reflecting the new opportunities of online and mobile channels [8]. Since 1991, Nielsen Soundscan began tracking sales data from cash registers collected from 14,000 retail, mass merchant, and non-traditional (on-line stores, venues, digital music services, etc.) outlets in the United States, Canada and the U.K. Previously, Billboard tracked sales by calling stores across the U.S. and asking about sales - a method that was inherently error-prone and open to outright fraud [6,10].

Traditionally, the record charts are based on weekly record sales. According to the IFPI (International Federation of the Phonographic Industry), the world's largest phonographic markets are: the USA, Japan, Great Britain, France and Germany [6,8]. 80% of weekly record sales belongs to the four biggest record companies (Universal, EMI, Sony BMG and Warner Bros). All the world's most popular artists are signed to these companies [6]. Thus, since 2003 it is possible to find their weekly record sales exactly [10].

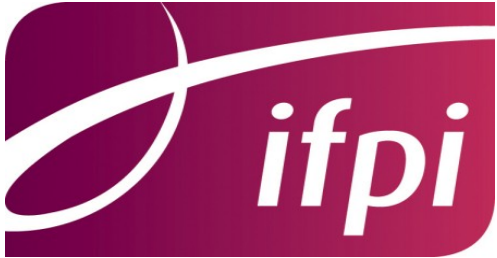

**Pic. 2** IFPI (International Federation of the Phonographic Industry)

The phonographic market differs from financial markets because price of a record (LP, CD, mp3) is constant. Therefore, the value of an artist is defined by weekly record sales. In my research, I have chosen the portfolio of the world's most popular 30 artists according to the record sales between May 2003 and November 2010.

The motivation of the present study concern the search for the kind of topological arrangement between artists and built Minimal Spanning Tree and the associated subdominant ultrametric hierarchical tree. This method has been successful in spin glass systems [11-12] and financial markets [13-15] because of meaningful economic taxonomy. The topology and hierarchical structure associated to it, has been obtained by using information present in the time series of stock prices only.

In this paper, I use these methods on phonographic market where we have artists instead of stocks. The value of an artist is defined by record sales. The graph is obtained starting from the matrix of correlations coefficient computed between the world's most popular 30 artists by considering the synchronous time evolution of the difference of the logarithm of weekly record sales.

The logarithm of weekly record sales or the logarithm of stock returns is used because of Fisher effect [16]. Thus, it ts possible to avoid influences of inflation in financial and phonographic markets. Logarithms were taken to make the series more nearly stationary.

## 2. Method of analysis: phonographic vs financial market

Criterion of the comparison given in Table 1 describes the analogy between financial and phonographic markets.

**Table 1**. Analogy between financial and phonographic markets

| Financial market | Phonographic market |
|---|---|
| stock prices | weekly record sales |
| price returns | change of record sales |
| correlation between stock prices | correlation between artists |
| distance between stocks | distance between artists |
| life time of correlations between stocks | life time of correlations between artists |
| main indice portfolio | top selling artists |
| industry sectors and subsectors | music genres |

The correlation coefficient defines a degree of similarity between the synchronous time evolution of a pair of assets:

$$\rho_{ij} = \frac{\langle Y_i Y_j \rangle - \langle Y_i \rangle \langle Y_j \rangle}{\sqrt{(\langle Y_i^2 \rangle - \langle Y_i \rangle^2)(\langle Y_j^2 \rangle - \langle Y_j \rangle^2)}} \qquad (1)$$

i and j are the numerical labels of assets, Yij is the price return. Yi = ln[Pi(t)] - ln[Pi(t – 1)] where Pi(t) is the weekly record sales of the artist i at the day t. The statistical average is a temporal average performed on all the trading days of investigated time period. By definition, ρij may vary from -1 to 1. The matrix of correlation coefficients is a symmetric matrix with ρii and the n(n - 1)/2 correlation coefficients characterize the matrix completely.

The correlation coefficient reflects similarity between assets. It can be used in building the hierarchical structure in financial markets and finding the taxonomy that allows to isolate groups of assets that make sense from an economic point of view [11]. In this research, matrix of correlations coefficient is computed between the world's most popular 30 artists by considering the synchronous time evolution of the difference of the logarithm of weekly record sales. The investigated time period is 1.05.2003 to 15.11.2010. I also computed correlation coefficients by considering the synchronous time evolution of the difference of the logarithm stock returns in this period to compare the markets. The results are given in Table 2.

Like in the financial market, three levels of correlations, given by (1), can be introduced :

1. Strong (strongly correlated pair of assets)   ρ c [1/2, 1] ;
2. Weak (weakly correlated pair of assets)   ρ c [0, 1/2) ;
3. Negative (anti-correlated pair of assets)   ρ c [-1, 0) .

This division is useful in finding the life-time of correlations between stocks in financial markets where almost all correlation coefficients are positive. It has been discussed in the previous chapter.

**Table 2**. Number of strongly, weakly and negatively correlated pairs in portfolios

| Correlated pairs | strongly | weakly | negatively |
|---|---|---|---|
| DJIA | 9 | 426 | 0 |
| DAX | 205 | 119 | 1 |
| WIG 20 | 1 | 188 | 1 |
| Phonographic market | 5 | 72 | 373 |

However, most of the correlation coefficients in the phonographic markets are negative (see Table 2) when artists essentially compete over the same group of customers. Such phenomena has been observed also in financial markets [17] - the success for one company often implies the failure for the others, indicating the market's reaction to the current situation (Futhermore, there is a fundamental negative correlation between gold-related stocks and the rest, indicating the complementary characters of these assets).

The correlation matrix can also be used in order to classify the artists into clusters. Like in the financial markets, the distance between artists is defined by:

$$d(i,j) = \sqrt{2\left(1-\rho_{ij}\right)} \qquad (2.2)$$

With this choice, d(i,j) fulfills three axioms of an Euclidean metric:

(i)  d **ij** = 0                          if and only if    i = j
(ii) d **ij** = d **ji**
(iii) d **ij** < d **ik** + d **kj**

## 2. Discussion and results

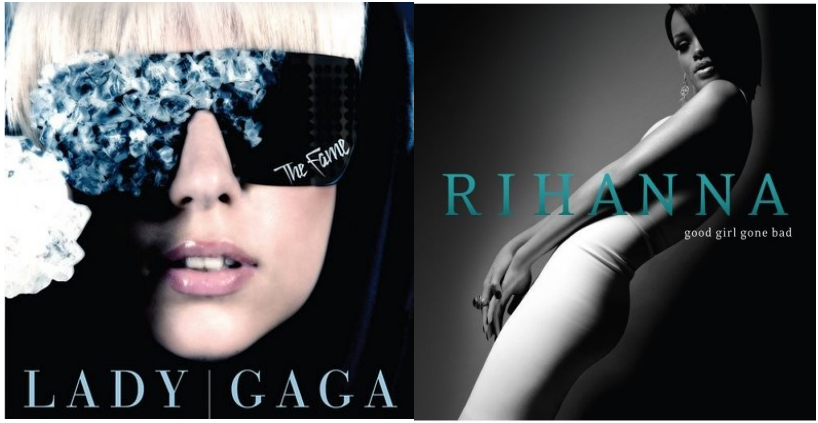

The strongest correlated pairs in phonographic market are:

| | |
|---|---|
| 0.73 | Lady Gaga – Rihanna (d = 0.73) |
| 0.69 | The Beatles – Jay-Z (d = 0.78) |
| 0.61 | The Beatles – U2 (d = 0.88) |
| 0.58 | Kanye West – 50 Cent (d = 0.91) |
| 0.54 | Jay-Z – U2 (d = 0.93) |

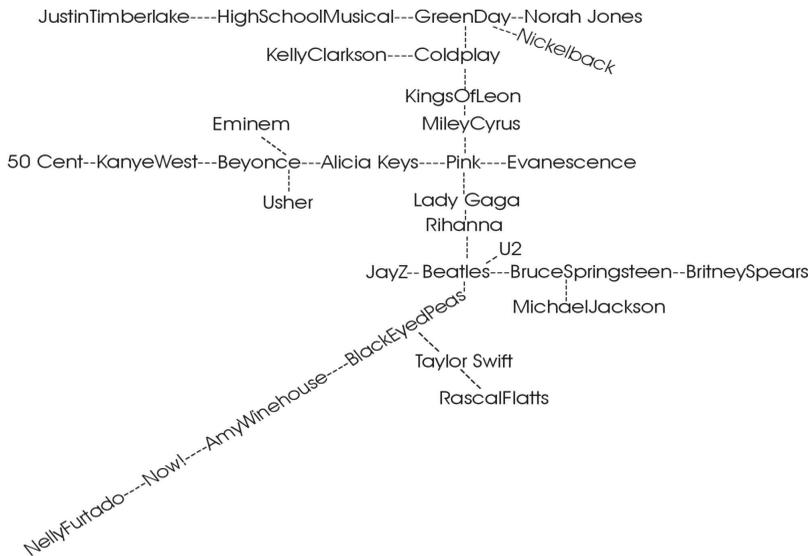

**Fig. 1.** The Minimum Spanning Tree connecting the world's most popular artists.

For that, According to [12] I have built the Minimal Spanning Tree (MST) for portfolio of stocks and portfolio of artists (Fig. 1). It provides an arrangement of assets, which selects the most relevant connections of each point of the set. The hierarchical structure in phonographic market is given in Fig. 2.

Although the hierarchical structure in financial markets (Fig 3) reflects the classifications of stocks in the industry sectors and sub-sectors [14-16] reported in the Forbes annual reports on American Industry, the analysis of the MST and correlations between artists does not always fit to music genres classified by the Billboard. The Minimal Spanning Trees reveals sectors that belong to rap (Eminem, 50 Cent, Kanye West), rock (Kings Of Leon, Coldplay, Kelly Clarkson, Miley Cyrus, Nickleback, Green Day), soul (Alicia Keys, Usher, Beyonce) or country artists (Rascal Flatts, Taylor Swift), but does not show the sector for pure pop music.

Instead of pop, we have celebrity sector that contain Jay-Z, The Beatles, Lady Gaga, Rihanna, U2, Michael Jackson, Bruce Springsteen and Britney Spears. What do they have in common? Although they represent various styles and genres, the only common thing they have is fame, high record sales, popularity and a place in music history. Because most of them were popular even before 2003.

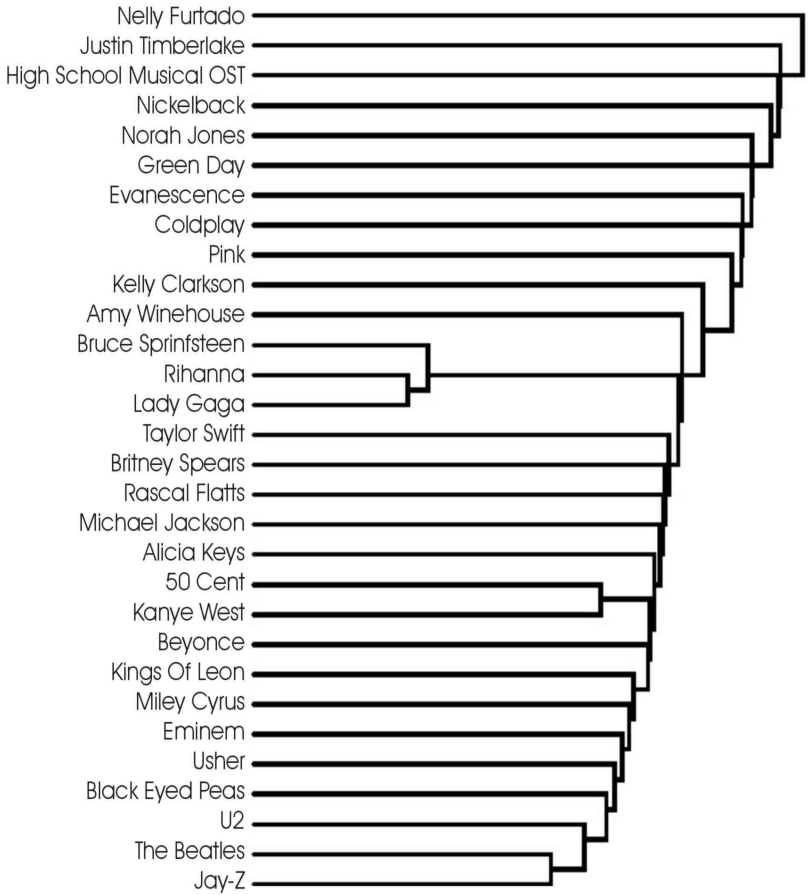

**Fig. 2.** Hierarchical structure in phonographic market

It is also interesting to check the durability of the Minimal Spanning Tree in phonographic market. For this purpose, I use tree half-life t1/2, defined as the time interval in which half the number of initial connections between artists have decayed [15]. The behavior of t1/2 as a function of the window width Δt is shown in Fig. 3. Like in financial markets, it follows a linear dependence for Δt < 1 yr after which it grows more than linear function. Such behavior can be affected by the product life-cycle, because artists release their next consecutive albums no more than once a year. For the linear region, the tree half-life exhibits t1/2 = 0,05 Δt dependence.

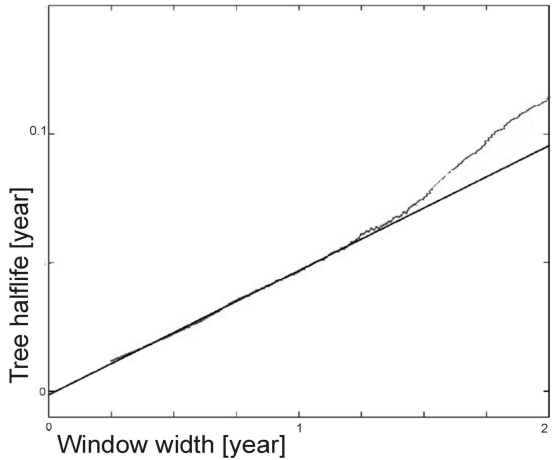

**Fig. 3**. Plot of a phonographic market tree half-life $t_{1/2}$ as a function of window width $\Delta t$

The conclusion is, that from economical point of view, pop music does not exist in the way that record companies think. Pop music is a term that originally derives from an abbreviation of "popular" and its extension to music genre is not allowed. The analysis of phonographic market based on the Minimal Spanning Tree revealed that there are two groups of customers. The first one buys music because of various genres, and the second one buys record because of an artists.

How record companies could use these results from the practical point of view? It is well known that the main goal of record companies is to increase record sales. Thus, they often insist on recording more 'pop' albums and are disappointed in record sales after release. Such strategy is wrong, because pop music is not well defined and not meaningful to customers. In practice, record companies should let artists develop their individuality or support the genres that have meaning from economic point of view.